\documentclass[10pt,twocolumn]{article}

\usepackage[margin=1in]{geometry}
\usepackage{times}
\usepackage{amsmath,amssymb}
\usepackage{graphicx}
\usepackage{booktabs}
\usepackage{url}
\usepackage{hyperref}
\usepackage{microtype}
\usepackage{xcolor}
\usepackage{enumitem}
\usepackage{caption}
\usepackage{subcaption}
\usepackage{algorithm}
\usepackage{algpseudocode}
\usepackage{multirow}
\usepackage{cite}
\usepackage{comment}

\usepackage{tikz}
\usetikzlibrary{arrows.meta, positioning, calc}
\pgfdeclarelayer{background}
\pgfsetlayers{background,main}

\definecolor{stageblue}{HTML}{6089BE}
\definecolor{stagered}{HTML}{C25B57}
\definecolor{stagegreen}{HTML}{9CBE62}
\definecolor{stagepurple}{HTML}{8571A9}
\definecolor{stageorange}{HTML}{F09146}
\definecolor{phasegrey}{HTML}{E8E8E8}

\def\StageData{%
    {Norm + QKV}/{Norm + QKV}/22/stageblue/{RMSNorm\\Q proj\\KV proj}/0.75,
    {Attention}/{Attention}/28/stagered/{RoPE Q/K\\FMHA}/1.0,
    {Output Proj /\\Residual}/{Output/Residual}/20/stageorange/{Output proj\\Add+Norm}/0.55,
    {Router /\\Dispatch}/{Router/Dispatch}/16/stagegreen/{Router gating\\TopK dispatch}/0.4,
    {Expert MLP /\\Combine}/{Expert/Combine}/62/stagepurple/{Expert up (BMM)\\SwiGLU\\Expert down\\Combine}/0.75%
}

\xdef\NumStages{0}\xdef\TotalSMs{0}
\foreach \t/\short/\sms/\col/\ops/\dur [count=\i] in \StageData {%
    \xdef\NumStages{\i}%
    \pgfmathtruncatemacro{\tmp}{\TotalSMs+\sms}\xdef\TotalSMs{\tmp}%
}

\def\SMScale{0.14}        % cm of box width per SM
\def\StageGap{0.55}       % horizontal gap between stage boxes (cm)
\def\StageHeight{4}     % stage box height (cm)
\def\TitleDrop{1.05}      % title distance below box top (cm)
\def\OpsDrop{0.40}        % op-list distance below box center (cm)

\def\PartitionSubtitle{Total = \TotalSMs{} SMs on one GPU (B200, $N_{SM}$=\TotalSMs)}
\def\FlowLabel{microbatch flow $\rightarrow$}

\def\NumMB{6}             % number of microbatches / iterations
\def\BarW{0.55}           % temporal panel: bar width (cm)
\def\BarH{1.9}            % temporal panel: bar height (cm)
\def\SlotW{0.85}          % SubGPU panel: time-slot width (cm)
\def\RowPitch{0.75}       % SubGPU panel: vertical distance between rows (cm)
\def\BlockH{0.62}         % SubGPU panel: block height (cm)
\def\PhaseStrip{0.7}      % extra grey strip below rows for phase labels (cm)

\tikzset{
    stage/.style={rounded corners=6pt, draw=black!70, line width=0.9pt,
                  minimum height=\StageHeight cm, text=white, align=center,
                  anchor=south west},
    stagetitle/.style={text=white, align=center, font=\sffamily\bfseries\large},
    oplist/.style={text=white, align=center, font=\sffamily},
    smlabel/.style={font=\sffamily\large},
    flow/.style={-{Stealth[length=3mm,width=3mm]}, line width=1.4pt, black!85},
    paneltitle/.style={anchor=west, font=\sffamily\bfseries\Large},
    annot/.style={font=\sffamily\itshape},
    phaselabel/.style={font=\sffamily\itshape},
    legenditem/.style={anchor=west, font=\sffamily, inner sep=2pt}
}

\newcommand{\DrawStagePartition}{%
    \xdef\StageCursor{0}%
    \foreach \t/\short/\sms/\col/\ops/\dur [count=\i] in \StageData {%
        \pgfmathsetmacro{\w}{\sms*\SMScale}%
        \node[stage, fill=\col, minimum width=\w cm] (s\i) at (\StageCursor,0) {};
        \node[stagetitle] at ($(s\i.north)+(0,-\TitleDrop)$) {\t};
        \node[oplist]     at ($(s\i.center)+(0,-\OpsDrop)$) {\ops};
        \node[smlabel, above=3pt of s\i.north] {\sms{} SMs};
        \pgfmathsetmacro{\nextx}{\StageCursor+\w+\StageGap}%
        \xdef\StageCursor{\nextx}%
    }%
    \pgfmathtruncatemacro{\lastpair}{\NumStages-1}%
    \foreach \i [evaluate=\i as \j using int(\i+1)] in {1,...,\lastpair}
        \draw[flow] (s\i.east) -- (s\j.west);
    \pgfmathsetmacro{\RightEdge}{\StageCursor-\StageGap}%
    \node[annot, font=\sffamily\itshape\Large]
        at (\RightEdge/2, \StageHeight+1.5) {\PartitionSubtitle};
    \node[anchor=east, font=\sffamily\bfseries\Large]
        at (\RightEdge+0.6, -0.5) {\FlowLabel};
}

\newcommand{\DrawTemporalPanel}{%
    \xdef\TCursor{0}%
    \foreach \m in {1,...,\NumMB} {%
        \foreach \t/\short/\sms/\col/\ops/\dur [count=\s] in \StageData {%
            \pgfmathsetmacro{\w}{\dur*\BarW*1.35}%
            \pgfmathsetmacro{\xb}{\TCursor+\w}%
            \fill[\col, draw=white, line width=0.6pt]
                (\TCursor, 0) rectangle (\xb, \BarH);
            \xdef\TCursor{\xb}%
        }%
    }%
    \begin{scope}[yshift=- 0.5cm]
    \foreach \t/\short/\sms/\col/\ops/\dur [count=\i] in \StageData {%
        \ifnum\i=1
            \node[legenditem] (leg\i) at (0,0)
                {\tikz\fill[\col] (0,0) rectangle (0.55,0.28); \short};
        \else
            \pgfmathtruncatemacro{\prev}{\i-1}%
            \node[legenditem] (leg\i) at ($(leg\prev.east)+(0.3,0)$)
                {\tikz\fill[\col] (0,0) rectangle (0.55,0.28); \short};
        \fi
    }%
\end{scope}
}

\newcommand{\DrawSubGPUPanel}{%
    \pgfmathsetmacro{\BandBottom}{-(\NumStages-1)*\RowPitch-\BlockH-\PhaseStrip}%
    \pgfmathtruncatemacro{\LastRow}{\NumStages-1}%
    \foreach \r in {0,...,\LastRow}{%
        \expandafter\xdef\csname FendRow\r\endcsname{0}%
    }%
    \foreach \m in {1,...,\NumMB} {%
        \xdef\PrevSegmentF{0}%
        \foreach \t/\short/\sms/\col/\ops/\dur [count=\r from 0] in \StageData {%
            \pgfmathsetmacro{\stagefree}{\csname FendRow\r\endcsname}%
            \pgfmathsetmacro{\start}{max(\PrevSegmentF,\stagefree)}%
            \pgfmathsetmacro{\fin}{\start+\dur*\SlotW}%
            \pgfmathsetmacro{\ya}{-\r*\RowPitch}%
            \fill[\col, draw=white, line width=0.6pt]
                (\start, \ya) rectangle (\fin, {\ya-\BlockH});
            \expandafter\xdef\csname FendRow\r\endcsname{\fin}%
            \xdef\PrevSegmentF{\fin}%
            \ifnum\m=1 \ifnum\r=\LastRow \xdef\FillEndX{\start}\fi\fi
            \ifnum\m=\NumMB
                \ifnum\r=0 \xdef\DrainStartX{\fin}\fi
                \ifnum\r=\LastRow \xdef\TotalEndX{\fin}\fi
            \fi
            \ifnum\m=1
                \node[anchor=east] at (-0.25, {\ya-\BlockH/2}) {\short};
            \fi
        }%
    }%
    \begin{pgfonlayer}{background}
        \fill[phasegrey] (0, \BandBottom) rectangle (\FillEndX, 0.55);
        \fill[phasegrey] (\DrainStartX, \BandBottom) rectangle (\TotalEndX, 0.55);
    \end{pgfonlayer}
    \node[phaselabel] at (\FillEndX/2, \BandBottom+0.3) {fill};
    \node[phaselabel, font=\sffamily\bfseries\itshape]
        at ({(\FillEndX+\DrainStartX)/2}, \BandBottom+0.3) {overlap};
    \node[phaselabel] at ({(\DrainStartX+\TotalEndX)/2}, \BandBottom+0.3) {drain};
}

\hypersetup{
  colorlinks=true,
  linkcolor=blue!60!black,
  citecolor=blue!60!black,
  urlcolor=blue!60!black,
  pdftitle={SubGPU Pipelining: Intra-GPU Fine-Grained Spatial Parallelism for LLM Inference},
  pdfauthor={Muralidhar Andoorveedu, Marius Juston, Sanjay Patel},
  pdfsubject={Technical Report -- Concept Preview, Version 1.0},
  pdfkeywords={LLM inference, intra-GPU parallelism, spatial scheduling, microbatching},
}

\usepackage{fancyhdr}
\title{\textbf{SubGPU Pipelining: Intra-GPU Fine-Grained\\
Spatial Parallelism for LLM Inference}
}

\author{
  Muralidhar Andoorveedu \quad Marius Juston \quad Tingkai Liu \quad Sanjay Patel \\
  NVIDIA \\
  \texttt{\{mandoorveedu, mjuston, tingkail, sanjpatel\}@nvidia.com}
}

\date{November 2026}

\begin{document}

\maketitle

% ============================================================
\begin{abstract}
Large language model (LLM) decode is dominated by short, predominantly memory-bound kernels that make bursty use of on-chip execution resources and memory bandwidth.
This paper introduces \emph{SubGPU Pipelining}, a fine-grained, intra-GPU spatial execution model that cuts an operator graph into ordered contiguous segments, binds those segments to spatial stages backed by disjoint streaming-multiprocessor (SM) pools, and streams microbatches through the resulting schedule. Selecting the graph cuts, segment-to-stage binding, per-stage SM allocation, and microbatch size is a coupled optimization problem: each choice changes the overlap pattern, shared-resource contention, cache state, and execution times of the others.
% Previous simulation scope:
% We evaluate the resulting design space with an analytical shared-resource simulator on Qwen3-235B-A22B-NVFP4 at sequence length 131{,}072 and deployments of up to 64 R100 GPUs. The model predicts up to $1.9\times$ higher per-user throughput at matched per-GPU throughput than temporal execution in the latency-sensitive decode regime.

We evaluate SubGPU Pipelining through both simulation and silicon experiment using the GPT-OSS 120B workload with 32-GPU B200 deployment settings. The analytical shared-resource simulator searches across
SubGPU pipeline structures and batch sizes to map the
per-user/per-GPU throughput Pareto frontier. At matched per-GPU throughput, the modeled SubGPU frontier provides up to $2.2\times$ higher per-user throughput than temporal execution, and at matched per-user throughput it provides up to $1.9\times$ higher per-GPU throughput.
To test physical feasibility and further validate the simulation results, silicon experiments are performed, using one B200 GPU to emulate a 32-GPU B200 deployment. The experiment evaluates one selected two-stage pipeline
configuration at global batch 2,048 using CUDA Green Contexts and TensorRT-LLM kernels. Results show that SubGPU Pipelining improves mean DRAM read-plus-write throughput from 71.8\% to 83.6\% of peak, and reduces the 36-layer decoder iteration time by 10.33\%.

% Previous prototype summary:
% To test physical feasibility, we built a minimal two-stage, three-segment prototype inside TensorRT-LLM using CUDA Green Contexts requiring zero changes to existing CUDA kernels.
% The prototype demonstrates spatial co-scheduling on a Blackwell GPU:
% mean utilized DRAM bandwidth rises from 69\% to 78\% of peak, and the
% mean layer period falls from $426.6\,\mu\mathrm{s}$ to
% $389.2\,\mu\mathrm{s}$ (8.8\% improvement) for the evaluated configuration.
\end{abstract}

% \par\medskip
% \noindent\begin{minipage}{\columnwidth}
% \textbf{Report scope.}
% This technical report records the SubGPU execution model, optimization
% problem, and initial evidence as of November 2026. It is intended as a concept
% preview rather than a complete reproducibility artifact or final performance
% characterization.
% \end{minipage}
%
% ============================================================
\section{Introduction}
\label{sec:intro}

The decode phase of LLM inference is the major source of cost in latency-sensitive serving. A single decode step can execute dozens of GPU kernels per transformer layer---normalization, attention, projection, routing, expert computation, residual addition, and more---each with distinct compute, memory, and communication requirements. Modern GPUs such as the NVIDIA B200~\cite{nvidia2026blackwell} are powerful enough that individual decode kernels at small batch sizes complete in microseconds, making launch overhead, idle compute cycles, and wasted HBM bandwidth important effects that collectively leave significant performance on the table.

Standard approaches---tensor parallelism (TP), data parallelism (DP), expert parallelism (EP), context parallelism (CP), and pipeline parallelism (PP)---scale the inference problem across GPUs, and do not directly address the intra-GPU inefficiencies just described.
Specifically, three pathologies remain regardless of how many GPUs are added:
\begin{enumerate}[noitemsep,leftmargin=*]
  \item \textbf{Kernel launch overhead.}  At small decode batch sizes, the dispatch latency for each kernel represents a non-trivial fraction of total step time.
  \item \textbf{Under-utilized compute.}  Memory-bound operators (\emph{e.g.}, expert projections at low-batch sizes) can saturate the HBM bandwidth roofline before consuming all SMs. Using all the SMs in such a case can mean the GPU is underutilized.
  \item \textbf{Bursty bandwidth demand.}  Alternating kernels with different bottlenecks create periods of under-utilization. For example even though the execution as a whole may be memory bound, compute-bound and latency-bound kernels during execution can cause memory bandwidth to be underutilized.
\end{enumerate}
%
% Previous prototype scope:
% This paper is a preview of a concept called \textbf{SubGPU Pipelining}. SubGPU Pipelining addresses inefficiency with LLM decode by cutting the temporal operator sequence within a layer (or, more generally, a repeating compute graph) into contiguous graph \emph{segments} and executing each segment concurrently on disjoint SM pools. The central problem is to jointly choose the graph cuts, segment-to-pool binding, and microbatch size so as to minimize target-workload makespan under shared-resource contention. We study this problem with simulation and provide an early proof-of-feasibility on a Blackwell GPU through a minimal prototype of a GPT-OSS 120B~\cite{openai2025gptoss} decode forward pass\footnote{Our GPT-OSS model only uses full attention, without sliding window attention}. A fuller end-to-end evaluation across model configurations and hardware generations is left to follow-on work.
This paper introduces a concept called \textbf{SubGPU Pipelining}. SubGPU Pipelining addresses inefficiency with LLM decode by cutting the temporal operator sequence within a layer (or, more generally, a repeating compute graph) into contiguous graph \emph{segments} and executing each segment concurrently on disjoint SM pools. The central problem is to jointly choose the graph cuts, segment-to-pool binding, and microbatch size so as to minimize target-workload makespan under shared-resource contention. We study this problem with simulation and provide proof-of-feasibility on a Blackwell GPU using the GPT-OSS 120B~\cite{openai2025gptoss} decode forward pass.

The key ideas are:
\begin{enumerate}[noitemsep,leftmargin=*]
  \item Cut a layer's operator graph into $K$ ordered contiguous segments and bind them to $D$ stages, each backed by a controlled, disjoint SM pool. $K=D$ or $K=D+1$; the latter represents the case where first and last segments may share a stage.
  \item Divide the per-GPU batch into $m$ microbatches and stream them through the segments while preserving graph dependencies and serializing work bound to the same stage.
  \item Jointly optimize graph cuts, stage count, segment-to-stage binding, per-stage SM allocation, and microbatch size because every segment's duration depends on the complete overlapping schedule.
\end{enumerate}
SubGPU Pipelining is conceptually composable with existing parallelism strategies: it operates within the per-GPU work allocation produced by an outer TP, DP, EP, CP, or PP configuration, provided that communication and shared-resource interactions are included in the schedule. The logical execution model does not prescribe a particular CUDA mechanism. 
% Our two-stage prototype binds three contiguous segments to two SM pools using existing kernels, while more aggressive implementations may use deeper pipelining, persistent kernels, or finer-grained signaling.

\paragraph{Contributions}
\begin{itemize}[noitemsep,leftmargin=*]
  \item A formal definition of SubGPU as a coupled graph segmentation, resource allocation, and microbatching problem with state-dependent segment times (\S\ref{sec:model}).
  \item A search and performance-modeling methodology for evaluating candidate SubGPU configurations (\S\ref{sec:search}).
  % Previous simulation contribution:
  % \item Analytical simulation on Qwen3-235B-A22B-NVFP4 that estimates the opportunity from spatial co-scheduling under explicitly stated modeling assumptions (\S\ref{sec:simulation-results}).
  \item Analytical simulation of GPT-OSS 120B on modeled B200
  deployments that estimates the opportunity from spatial co-scheduling under
  explicitly stated modeling assumptions (\S\ref{sec:simulation-results}).
  % Previous prototype contribution:
  % \item A minimal two-stage, three-segment Blackwell prototype in TensorRT-LLM---requiring zero changes to existing CUDA kernels---that demonstrates coarse spatial co-scheduling, raises measured average DRAM throughput, and reduces layer time (\S\ref{sec:hardware-results}).
  \item A Blackwell prototype in TensorRT-LLM, requiring zero changes to existing CUDA kernels, that demonstrates coarse spatial co-scheduling, raises measured
  average DRAM throughput, and reduces iteration time
  (\S\ref{sec:hardware-results}).
\end{itemize}
%
% ============================================================
\section{Background and Related Work}
\label{sec:related}

\subsection{Pipeline Parallelism Across GPUs}

Pipeline Parallelism (PP) is a widely used technique for scaling deep
learning training and inference across multiple GPUs. GPipe~\cite{huang2019gpipe}
introduced microbatch pipeline parallelism, mapping disjoint sets of model
layers onto different devices and flowing microbatches through the resulting
pipeline. PipeDream~\cite{narayanan2019pipedream} uses asynchronous pipeline
schedules to reduce fill and drain bubbles, while Megatron-LM~\cite{narayanan2021megatron}
combines pipeline, tensor, and data parallelism and uses interleaved pipeline
schedules at large scale.

CTA-Pipelining~\cite{liu2026ctapipelining} moves the dependency granularity
below whole kernels to cooperative thread arrays (CTAs), also called CUDA
thread blocks, on tightly connected, shared-memory multi-GPU systems. It
allows producer and consumer CTAs from dependent kernels to make progress
concurrently on different GPUs and can compose with tensor parallelism. These
approaches distribute pipeline work across GPUs and do not directly address
intra-GPU resource under-utilization.

\subsection{Intra-Device Parallelism}

Earlier work proposed hardware, compiler, and runtime support for pipelines within a GPU.
Kim et al.~\cite{kim2016implicitpipeline} use reference-count scoreboarding and
a pipeline-aware CTA scheduler to overlap independent portions of consecutive kernels. VersaPipe~\cite{zheng2017versapipe}
includes persistent-thread execution models, explicit stage-to-SM mapping, and
an autotuned hybrid of coarse- and fine-grained GPU pipelines. cuSync retains
separate CUDA kernels but replaces kernel-wide barriers with tile-grained
producer--consumer synchronization, allowing independent tiles of dependent
kernels to overlap~\cite{jangda2024cusync}. Rammer lowers operators into
fine-grained tasks and constructs static spatio-temporal schedules that expose
both inter- and intra-operator parallelism~\cite{ma2020rammer}.

Several recent LLM systems concurrently use a GPU's compute, memory, and
communication resources.
NanoFlow~\cite{zhu2025nanoflow} splits serving batches into nano-batches and
jointly searches their sizes, execution order, and fractional GPU resource
allocations while accounting for concurrent interference.
Flux~\cite{chang2024flux} fuses collective communication with computation,
Comet~\cite{zhang2025comet} applies
fine-grained fused overlap to MoE workloads, and
TokenWeave~\cite{gond2025tokenweave} uses wave-aware token splitting to overlap
computation and communication. DynaFlow~\cite{dynaflow2026} decouples the
logical model from its physical schedule through a programmable frontend and
an asynchronous runtime.

\paragraph{Megakernels}
A complementary approach replaces a sequence of operator kernels with one
persistent megakernel. ThunderKittens provides CUDA-embedded abstractions for
constructing high-performance kernels~\cite{spector2025thunderkittens}; a
ThunderKittens-based Llama implementation fuses a complete forward pass,
assigns precomputed instruction streams to SMs, and tracks dependencies within
the kernel~\cite{spector2025megakernel}. MPK compiles tensor programs into
SM-level task graphs and executes them with decentralized scheduling inside a
persistent megakernel, enabling cross-operator software pipelining and
fine-grained overlap~\cite{cheng2026mpk}.

SubGPU focuses on a latency-oriented formulation over contiguous graph cuts,
segment-to-stage binding, disjoint SM allocations, and microbatch size. Its
logical model is independent of a particular dispatch mechanism;
Section~\ref{sec:realization} discusses persistent kernels and other CUDA
mechanisms as possible realizations and extensions.

\subsection{Spatial Dataflow Execution}

Kitsune augments a GPU with architectural primitives and compiler support for
co-resident spatial dataflow pipelines~\cite{davies2025kitsune}.
Purpose-built dataflow accelerators pursue a related spatial execution model
at the architecture level. SambaNova's RDU maps computation, local storage,
and data movement onto a reconfigurable fabric of compute units, memory units,
and programmable interconnect~\cite{prabhakar2022sn10}. Recent work on the
SN40 applies this substrate to LLM decoding by fusing large decoder subgraphs,
streaming intermediate values on chip, and overlapping samples across layers
through BatchStreaming~\cite{gandhi2026dataflow}.

\subsection{GPU Resource Partitioning}

SM-level spatial sharing predates recent CUDA mechanisms. Adriaens et
al.~\cite{adriaens2012spatial} partition SMs among concurrent GPU applications;
Janz\'en et al.~\cite{janzen2016partitioning} enforce disjoint SM subsets in
software and use partition-specific models to schedule a precedence-constrained
task graph; and Zhao et al.~\cite{zhao2018classification} search SM partitions
while accounting for bandwidth and cache interference.

CUDA Streams allow concurrent kernel submission but provide no guarantee of co-resident execution or an explicit SM partition. Cooperative Groups and Multi-Process Service provide other forms of coordination and co-scheduling. CUDA Green Contexts~\cite{cuda124relnotes} provide first-class SM partitioning within a single CUDA context: streams within a Green Context are bound to an SM subset, and kernels launched on those streams execute on that subset. Persistent kernels, CUDA events, and CUDA Graphs~\cite{nvidia2026cudaguide} provide different choices for dispatch and dependency signaling. These are possible realization mechanisms; the SubGPU execution model itself requires controlled resource partitioning and dependency-preserving progress, while useful performance also requires sufficiently low handoff overhead.

% ============================================================
\section{SubGPU Execution and Optimization Problem}
\label{sec:model}

\subsection{Execution Model and Design Space}

\begin{figure*}[t]
  \centering
  \includegraphics[width=\linewidth]{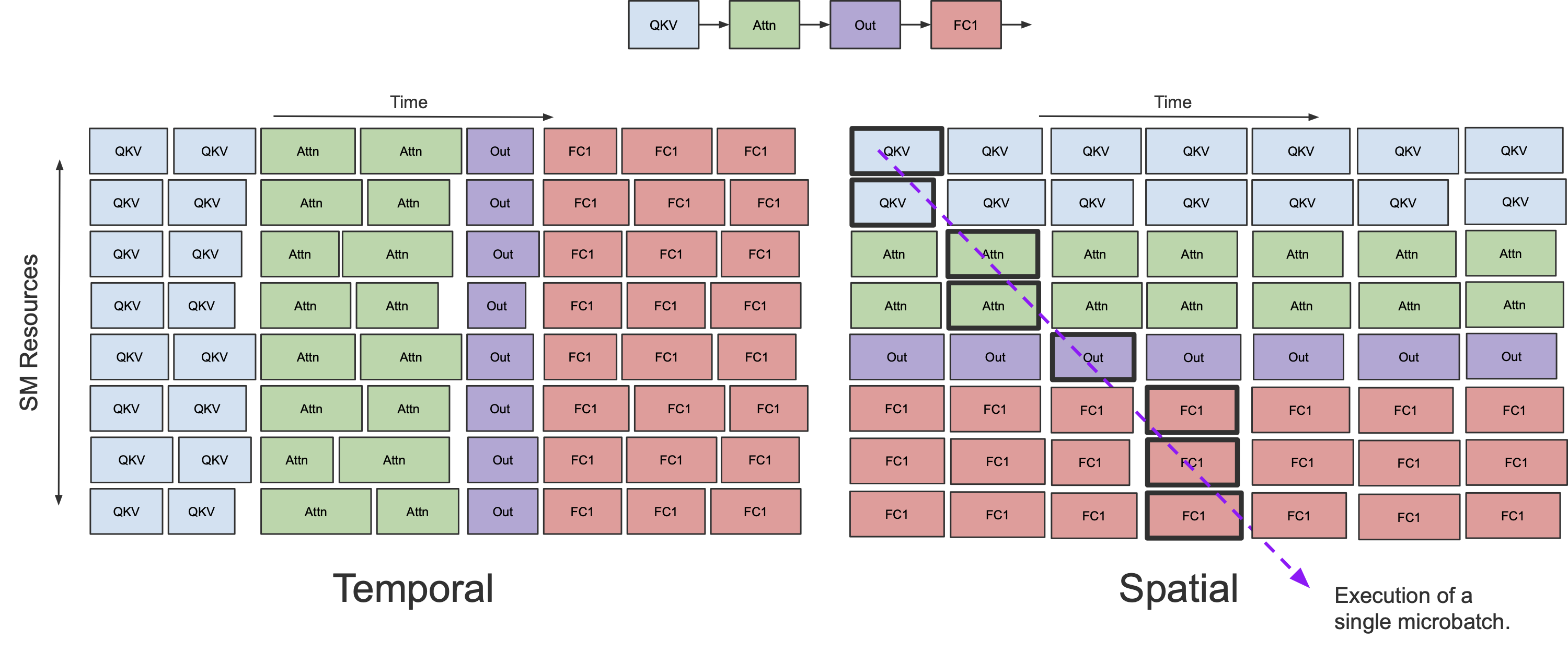}
  \caption{Temporal execution (left) versus an illustrative one-to-one SubGPU binding (right). Temporal execution executes one operator at a time, with the full GPU. SubGPU binds contiguous graph segments to disjoint SM pools and streams microbatches through them, allowing work with complementary resource demands to overlap. In this example, the temporal case is executing one batch across multiple waves.  The spatial case is executing the same batch across 8 microbatches.}
  \label{fig:temporal_vs_spatial}
\end{figure*}

Figure~\ref{fig:temporal_vs_spatial} contrasts conventional temporal execution with an illustrative SubGPU schedule. Let $G=(V,E)$ be the operator DAG for a repeating region, such as one
transformer layer. Here, $E$ contains only dependencies internal to the
region; cross-region dependencies are handled when regions are composed into
the full forward-pass DAG.  Each contiguous region of operators between two graph cuts is a
\emph{segment}, which will be provisioned a disjoint SM pool on which to execute.  From a scheduling perspective, the resource-mapped segment forms a \emph{stage} in the SubGPU pipeline. Let $\mathcal C_K(G)$ contain the legal ways to cut $G$
into $K$ nonempty ordered, dependency-contiguous segments
$c=(S_1,\ldots,S_K)$. Each prefix
$S_1\cup\cdots\cup S_k$ must be predecessor-closed, so every boundary is a
dependency-respecting graph cut.

A configuration uses $D$ physical stages and either $K=D$ or $K=D+1$
segments. The latter case captures the periodic boundary: the last segment
of one region and the first segment of the next may use the same stage.
Numbering stages by first appearance, let $w\in\{0,1\}$ indicate this
boundary reuse, so $K=D+w$/. For segment indices $k\in\{1,\ldots,K\}$, define the segment-to-stage
binding
\begin{equation}
  \rho_w(k)=
  \begin{cases}
    1, & w=1\ \text{and}\ k=K,\\
    k, & \text{otherwise}.
  \end{cases}
  \label{eq:segment-binding}
\end{equation}
Thus the binding is one-to-one when $w=0$; when $w=1$, only the first
and last segments share a stage. We represent a configuration as
\begin{equation}
  \theta=(D,w,c,p,b),
  \label{eq:configuration}
\end{equation}
where $p=(p_1,\ldots,p_D)$ gives the SM allocation of the $D$ physical
stages and $b$ is the microbatch size. Let $H$ denote the target GPU hardware,
including its SM count, $N_{SM}$ and minimum per-stage SM allocation, $p_{\min}$. The $D$ stage allocations correspond to pairwise-disjoint SM subsets, so no SM is assigned to more than one physical stage.
Let $\mathcal P_H^{(D)}$ contain the legal ordered $D$-stage SM-allocation vectors for $H$, with
$p_d\geq p_{\min}$ and
$\sum_d p_d = N_{\mathrm{SM}}$, and let $\mathcal B(B)$ contain the
supported microbatch sizes that divide the per-GPU batch $B$. Feasibility
is summarized by
\begin{equation}
  \begin{aligned}
    &w\in\{0,1\},\quad 2\leq D\leq |V|-w,\\
    &c\in\mathcal C_{D+w}(G),\\
    &p\in\mathcal P_H^{(D)},\quad b\in\mathcal B(B).
  \end{aligned}
  \label{eq:feasibility}
\end{equation}
Let $\mathcal F$ denote the configurations satisfying
Equation~\ref{eq:feasibility}.
The choice of $b$ gives $m=B/b\in\mathbb Z_{+}$ microbatches.
Smaller microbatches expose more independent work and can fill more stages,
but they may also reduce the efficiency of each kernel invocation; this is why
$b$ is optimized jointly with the graph cuts and SM allocation.

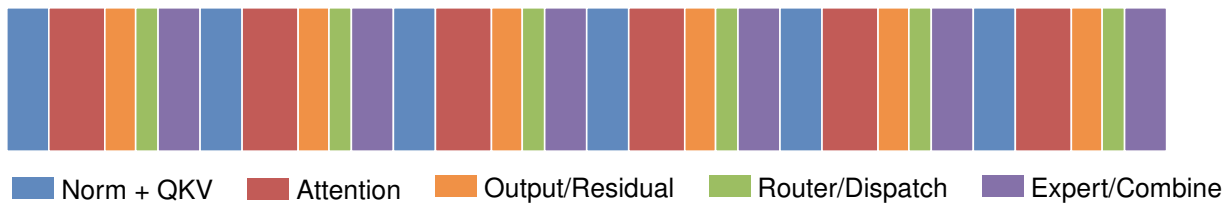
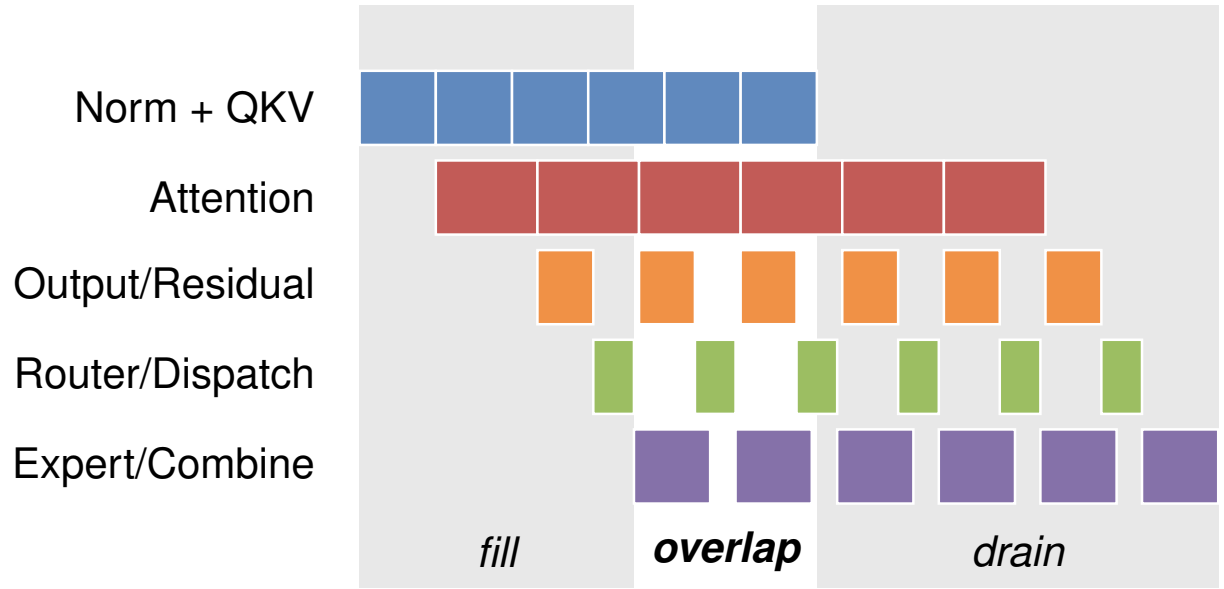
\begin{figure}[t]
  \centering
  \begin{subfigure}[b]{\linewidth}
    \centering
    \resizebox{\linewidth}{!}{%
      \begin{tikzpicture}[font=\sffamily]
        \DrawTemporalPanel
      \end{tikzpicture}%
    }
    \caption{Temporal baseline: segments execute serially on the full GPU.}
    \label{fig:pipeline-temporal}
  \end{subfigure}

  \vspace{0.6em}

  \begin{subfigure}[b]{\linewidth}
    \centering
    \resizebox{\linewidth}{!}{%
      \begin{tikzpicture}[font=\sffamily]
        \DrawSubGPUPanel
      \end{tikzpicture}%
    }
    \caption{SubGPU: segments on distinct stages overlap subject to data and
    same-stage dependencies.}
    \label{fig:pipeline-subGPU}
  \end{subfigure}

  \caption{Illustrative schedules for six microbatches under a chain-like,
  five-segment, one-to-one binding. Rectangle widths use hand-chosen relative
  durations to expose unequal task lengths and scheduling bubbles; they are not
  simulator outputs and do not imply that a segment has the same duration
  across microbatches or overlap states. Actual times follow
  Equation~\ref{eq:task-timing} using contention-dependent per-task durations
  supplied by the simulator. The panels are independently scaled and do not
  compare absolute makespan.}
  \label{fig:pipeline-comparison}
\end{figure}

Figure~\ref{fig:pipeline-comparison} gives scheduling intuition for the
chain-like, one-to-one case. Each segment is gated by both its data
dependencies and the preceding segment assigned to the same physical stage; the
next subsection formalizes these constraints.

\begin{figure*}[t]
  \centering
  \resizebox{\textwidth}{!}{%
    \begin{tikzpicture}[font=\sffamily]
      \DrawStagePartition
    \end{tikzpicture}%
  }
  \caption{Illustrative five-stage SubGPU configuration for one MoE transformer layer on a single B200 GPU. Five contiguous graph segments bind one-to-one to five SM pools; box width is proportional to SM allocation. This is one feasible configuration, not a prescribed segmentation. FMHA denotes fused multi-head attention, and BMM denotes batched matrix multiplication.}
  \label{fig:partition}
\end{figure*}
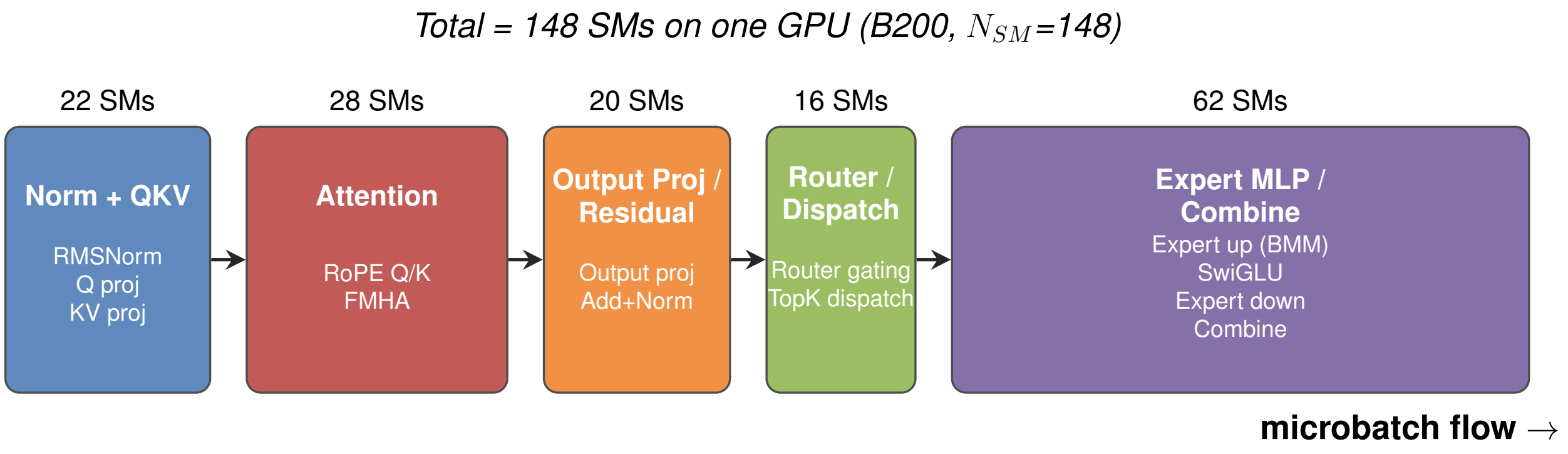

Figure~\ref{fig:partition} shows the one-to-one case: five contiguous
segments bind to five physical stages with different SM allocations.

\subsection{Augmented Task DAG and Objective}

For a configuration $\theta$, unrolling one forward pass over its region
instances and $m$ microbatches yields segment tasks $\mathcal J_\theta$.
A task $i=(\ell,\mu,k)$ executes segment
$S_k$ in region instance $\ell$ for microbatch $\mu$, on stage
$\rho_w(k)$. A region instance is one occurrence of the repeating graph $G$ in the
unrolled forward pass; for a transformer model, it typically corresponds
to one layer.

For task $i$, $P_i^{\mathrm{data}}$ contains the tasks whose outputs it
consumes, including cross-region dependencies. A fixed global dispatch order
is chosen as a topological ordering of these data dependencies. The set
$P_i^{\mathrm{stage}}$ is empty or contains the preceding task on the same
physical stage after that global order is restricted to the stage. These stage
edges serialize all work on a partition, including the boundary segments that
share stage 1 when $w=1$. Both edge types therefore point forward in the
same order and together define an acyclic augmented task DAG.

Intuitively, a task starts only after all required outputs have arrived and its
stage is free. It then occupies that stage for an effective duration $\tau_i$,
supplied by Section~\ref{sec:search}; the forward pass ends when the last task
finishes.

In the formal model, we assume dependency signaling and inter-segment handoff
latency are negligible, so data and stage edges impose precedence but no
separate delay.

\begin{samepage}
For any realized effective durations, let $A_i$ and $C_i$ denote start
and completion times. Topological evaluation then gives
\begin{equation}
  \begin{aligned}
    A_i &= \max\left\{
      \max_{j\in P_i^{\mathrm{data}}}C_j,
      \max_{j\in P_i^{\mathrm{stage}}}C_j
    \right\},\\
    C_i &= A_i+\tau_i,
  \end{aligned}
  \label{eq:task-timing}
\end{equation}
\end{samepage}
We take an empty maximum to be zero. The forward-pass makespan is
\begin{equation}
  T(\theta)=
  \max_{i\in\mathcal J_\theta} C_i.
  \label{eq:forward-makespan}
\end{equation}
For fixed $\tau_i$, Equations~\ref{eq:task-timing} and
\ref{eq:forward-makespan} are a max-plus timing recurrence on a weighted
event DAG and reduce to a longest-path calculation
~\cite{baccelli1992synchronization}. Section~\ref{sec:search} separately
describes how the simulator estimates $\tau_i$ under overlap and
shared-resource contention.

The optimization objective is
\begin{equation}
  \theta^\star \in
  \arg\min_{\theta\in\mathcal F}
  T(\theta).
  \label{eq:optimization-objective}
\end{equation}
The timing calculation is simple, but the optimization remains coupled:
graph cuts change the tasks and data edges, boundary reuse changes the stage
edges, and SM allocation and microbatching change the durations and critical
path.

The reported results use static configurations; Section~\ref{sec:realization}
discusses dynamic reconfiguration.

\subsection{Scale of the Search Space}
\label{sec:search-space}

An \emph{outer parallelism configuration} $q$ specifies the GPU count and
the applicable TP, DP, EP, CP, and PP degrees used to distribute the model. It
induces the per-GPU operator graph $G_q$, local tensor shapes, and collective
communication tasks. Let $\mathcal X$ be the finite set of feasible search points
$x=(q,B,b,D,\pi)$ for a given model, hardware, sequence length, and
repeating-region definition, where $B$ and $b$ are the per-GPU batch and
microbatch, $D$ is the physical-stage count, and $\pi$ is the fixed
simulator-policy setting. Let $\mathcal P_x$ contain the legal SM allocations
at $x$. An exhaustive sweep would contain
\begin{equation}
  N_{\mathrm{raw}}=
  \sum_{x\in\mathcal X}
  |\mathcal P_x|
  \left(
    |\mathcal C_D(G_q)|+|\mathcal C_{D+1}(G_q)|
  \right),
  \label{eq:search-space-size}
\end{equation}
where the two cut terms count the one-to-one and boundary-reuse cases; an
invalid cut set is empty. At one illustrative grid point, a 20-operator chain
with five stages has $15{,}504$ such cut-and-binding choices. Under the
reported full-budget, two-SM-quantum policy, dividing 148 SMs among five
nonempty ordered stages gives $1{,}088{,}430$ allocations---about
$1.7\times10^{10}$ combinations before counting any other point in
$\mathcal X$.

Brute-force evaluation therefore quickly becomes impractical as the outer
grid grows. More importantly, changing a cut, allocation, or microbatch changes
overlap, shared-resource contention, cache behavior, and completion times, so
candidates cannot be scored by adding isolated kernel costs.
Section~\ref{sec:search} therefore uses inexpensive proposals and refinement
to explore a small part of the space, and the event-driven simulator to
evaluate the coupled schedules that remain.

% ============================================================
\section{Search and Performance Model}
\label{sec:search}

We now describe the event-driven model used to evaluate candidates and the
staged procedure used to search the space in Section~\ref{sec:search-space}.
Lower-level implementation constants and policies are outside the scope of
this design report.

\subsection{Event-Driven Simulator}

The augmented task DAG fixes data precedence and the FIFO order on each physical
stage. Internally, the simulator expands every segment task into its constituent
kernel tasks. A stage executes only its head task and cannot bypass blocked
work. Whenever a dependency releases a task or an active work counter reaches
zero, the simulator recomputes completion times for the active set and advances
to the next event. This captures fill, steady state, drain, recurrence across
regions, and head-of-line blocking without assuming constant or balanced stage
times. The resulting start and completion times determine the effective
duration $\tau_i=C_i-A_i$ used by Section~\ref{sec:model}.

Each kernel task carries analytical compute and memory work together with a
modeled full-GPU compute rate. The model assumes that compute throughput
scales linearly with the number of SMs assigned to the task's stage:
\begin{equation}
  r_u^{\mathrm{comp}}=
  \frac{p_{d(u)}}{N_{\mathrm{SM}}}\mu_u,
  \label{eq:compute-scaling}
\end{equation}
where $u$ is a kernel task, $d(u)$ is its physical stage, and $\mu_u$
is its modeled full-GPU rate. Exposed memory concurrency grows with the
assigned SM count but is capped by a calibrated per-SM estimate, a hardware
limit, and the task's remaining demand. L2-hit and HBM-bound traffic jointly
consume L2 bandwidth; aggregate miss-side concurrency selects a nonlinear HBM
bandwidth response; and each task is limited by the issue bandwidth of its
assigned SMs. Concurrent tasks also share network bandwidth. Compute and data
movement may progress concurrently, and a task completes only after all of its
modeled work is finished. The same fixed sharing and cache policies are used
for Temporal and SubGPU candidates within an evaluation sweep.

The cache model tracks capacity-bounded residency of logical weight and
activation objects across stages, layers, and microbatches, charging misses to
HBM. The resident table begins empty for each simulation and evolves according
to deterministic admission and eviction policies. Activation writes may seed
later reads and evict resident objects. KV-cache accesses are modeled as
streaming traffic and are not admitted to the resident table. Weight reuse is
therefore an outcome of the simulated schedule rather than an assumed SubGPU
benefit. Collective operations are explicit tasks with setup latency and
payload-dependent fabric work, and compute kernels are configured with base launch overhead.
% Previous overhead model:
% set kernel-launch, dependency-signaling, and inter-segment handoff overhead to
% zero for both execution models while retaining communication setup latency.

\subsection{Search Procedure}

For each workload and deployment choice, the search instantiates the local operator graph and rejects
infeasible configurations. It then proposes a graph segmentation, refines the
SM allocation, and evaluates the resulting candidate. Prior work likewise uses performance models or profiling to account for co-scheduled interference while searching SM partitions and intra-device
schedules~\cite{zhao2018classification,zhu2025nanoflow}.  The search is expensive enough that it is best to run offline, with pre-computed configurations selected by the inference engine.

\begin{algorithm}[!t]
  \caption{Simulation-guided SubGPU search}
  \label{alg:subGPU-search}
  \footnotesize
  \begin{algorithmic}[1]
      \Require base DAG $G$, hardware $H$, outer grid $\mathcal X$
      \Ensure reported nondominated configurations
      \Statex \textbf{Grid point:}
        $x=(q,B,b,D,\pi)$ specifies outer parallelism configuration, per-GPU batch,
        microbatch, physical-stage count, and fixed sharing/cache policies.
      \Statex \textbf{Derived objects:}
        $G_q$ is the instantiated local DAG; $\phi_x$ contains kernel
        resource features; $\mathcal P_x$ is the set of legal SM allocations.
      \Statex \textbf{Partition:}
        $c$ gives the contiguous cuts and $w$ indicates boundary reuse.
      \Statex \textbf{Objectives:}
        for fixed $(x,c,w)$, $\widehat f_x(p)$ estimates makespan and
        $f_x(p)$ obtains it from the event simulator.
      \Statex \textbf{Controls:}
        $g=2$ SMs, $L_{\mathrm{est}}\in\mathbb N\cup\{\infty\}$, and
        $L_{\mathrm{sim}}\in\mathbb N$; the default uses
        $L_{\mathrm{est}}=\infty$.
      \State $\mathcal Q\gets\varnothing$
      \ForAll{$x=(q,B,b,D,\pi)\in\mathcal X$}
        \State $G_q\gets\Call{Instantiate}{G,q}$
        \If{$\neg\Call{Feasible}{x,G_q,H}$}
          \State \textbf{continue}
        \EndIf
        \State $\phi_x\gets\Call{KernelFeatures}{G_q,x,H}$
        \State $(c,w)\gets\Call{ProposePartition}{G_q,D,\phi_x}$
        \If{no legal partition is returned}
          \State \textbf{continue}
        \EndIf
        \State $\mathcal S\gets
          \Call{InitialAllocations}{c,w,\phi_x,\mathcal P_x}$
        \If{$\mathcal S=\varnothing$} \State \textbf{continue} \EndIf
        \State $\mathcal C_x\gets\varnothing$
        \ForAll{$p_0\in\mathcal S$}
          \State $\widehat p\gets
            \Call{PairwiseLine}{\widehat f_x,\mathcal P_x,p_0,g,L_{\mathrm{est}}}$
          \ForAll{$s\in\{p_0,\widehat p\}$}
            \State $p_s\gets
              \Call{PairwiseLine}{f_x,\mathcal P_x,s,g,L_{\mathrm{sim}}}$
            \State $\mathcal C_x\gets\mathcal C_x\cup\{p_s\}$
          \EndFor
        \EndFor
        \State $p^\star\gets\arg\min_{p\in\mathcal C_x}f_x(p)$
        \State $p^\star\gets
          \Call{PairwiseLine}{f_x,\mathcal P_x,p^\star,g,L_{\mathrm{sim}}}$
        \State $\mathcal Q\gets\mathcal Q\cup
          \{\Call{Record}{x,c,w,p^\star,f_x(p^\star)}\}$
      \EndFor
      \State \Return $\Call{Pareto}{\mathcal Q}$
  \end{algorithmic}
\end{algorithm}

\begin{algorithm}[!t]
  \caption{Pairwise-line SM-allocation refinement}
  \label{alg:pairwise-line}
  \footnotesize
  \begin{algorithmic}[1]
      \Require objective $f$, feasible set $\mathcal P$,
        initial allocation $p\in\mathcal P$, SM quantum $g$,
        pass limit $L\in\mathbb N\cup\{\infty\}$
      \Ensure feasible refined allocation $p$
      \Statex \textbf{Move:}
        $p^{d\rightarrow r}[z]$ moves $zg$ SMs from donor stage $d$
        to receiver stage $r$, leaving every other allocation unchanged.
      \Statex \textbf{Local variables:}
        $\mathcal M$ lists legal one-quantum transfers; $t$ is the current
        objective; $z$ counts transferred quanta; $h$ is the line-search
        step; and $t_{\mathrm{line}}$ is the best value found on that line.
      \Statex If $L=\infty$, refinement continues until an early return.
      \State $t\gets f(p)$
      \For{$\mathit{pass}=1,\ldots,L$}
        \State $\mathcal M\gets
          \{(d,r):d\neq r,\ p^{d\rightarrow r}[1]\in\mathcal P\}$
        \If{$\mathcal M=\varnothing$} \State \Return $p$ \EndIf
        \State $(d,r)\gets
          \arg\min_{(d',r')\in\mathcal M}f(p^{d'\rightarrow r'}[1])$
        \State $z\gets1;\ h\gets1;\
          t_{\mathrm{line}}\gets f(p^{d\rightarrow r}[1])$
        \If{$t_{\mathrm{line}}\geq t$} \State \Return $p$ \EndIf
        \While{$h>0$}
          \State $z'\gets z+h$
          \If{$p^{d\rightarrow r}[z']\in\mathcal P\ \land\
            f(p^{d\rightarrow r}[z'])<t_{\mathrm{line}}$}
            \State $z\gets z';\
              t_{\mathrm{line}}\gets f(p^{d\rightarrow r}[z']);\
              h\gets2h$
          \Else
            \State $h\gets\lfloor h/2\rfloor$
          \EndIf
        \EndWhile
        \State $p\gets p^{d\rightarrow r}[z];\
          t\gets t_{\mathrm{line}}$
      \EndFor
      \State \Return $p$
  \end{algorithmic}
\end{algorithm}

The partition proposal summarizes each kernel's expected compute, memory,
cache, and communication behavior using the simulator inputs described in algorithm \ref{alg:subGPU-search}.
Dynamic programming tries to find an anchored, dependency-respecting contiguous partition that minimizes differences between kernels in each segment. This is to promote the sharing of different resources as segments overlap. A separate step evaluates legal circular rotations; an
improving rotation may produce the boundary-reuse layout in
Equation~\ref{eq:segment-binding}. The feature objective proposes plausible
cuts rather than predicting execution time.

Uniform and service-proportional SM allocations seed the allocation search. For
each seed, a cheap dependency-aware makespan estimate first guides pairwise SM
transfers. The event simulator then polishes both the original allocation and
its estimator-refined counterpart, followed by a final polish of the best
candidate. Simulator evaluations are memoized. By default, estimator-guided
refinement continues until no improving transfer remains, while each
simulator-guided polishing call has a bounded pass count.

The reported search assigns each physical stage a positive, even number of SMs
and uses the full SM budget. It tests microbatch counts below, near, and above
the physical-stage count; Temporal uses one full-batch microbatch. These are
search choices rather than requirements of the SubGPU model. After the sweep,
the reported configurations are Pareto-filtered.
Algorithms~\ref{alg:subGPU-search} and~\ref{alg:pairwise-line} give the
corresponding procedures.

\subsection{Calibration and Scope}

Kernel work estimates come from analytical operator shapes and deployment
sharding. Hardware inputs provide datatype-specific compute throughput, HBM
and L2 capacity and bandwidth, per-SM issue bandwidth, communication latency
and bandwidth, and an empirical HBM bandwidth-versus-concurrency curve. A
calibrated per-kernel bytes-in-flight estimate and its global scale determine
how quickly each kernel exposes memory-level parallelism as SMs are added.
Collective schedules are selected from modeled latency/bandwidth costs before
global fabric contention is applied.

The outer sweep and feasible allocation sets are finite. Each allocation search
terminates when no feasible pairwise move improves the objective or when an
applicable pass limit is reached. Small graphs and SM budgets can be
exhaustively enumerated to diagnose the segmentation and allocation heuristics.
The reported points are therefore the best configurations found over the finite
search grid under these modeling assumptions, not proofs of global optimality.

% ============================================================

\begin{figure*}[ht]
  \begin{minipage}[t]{0.485\textwidth}
    \centering
    \includegraphics[width=\linewidth]{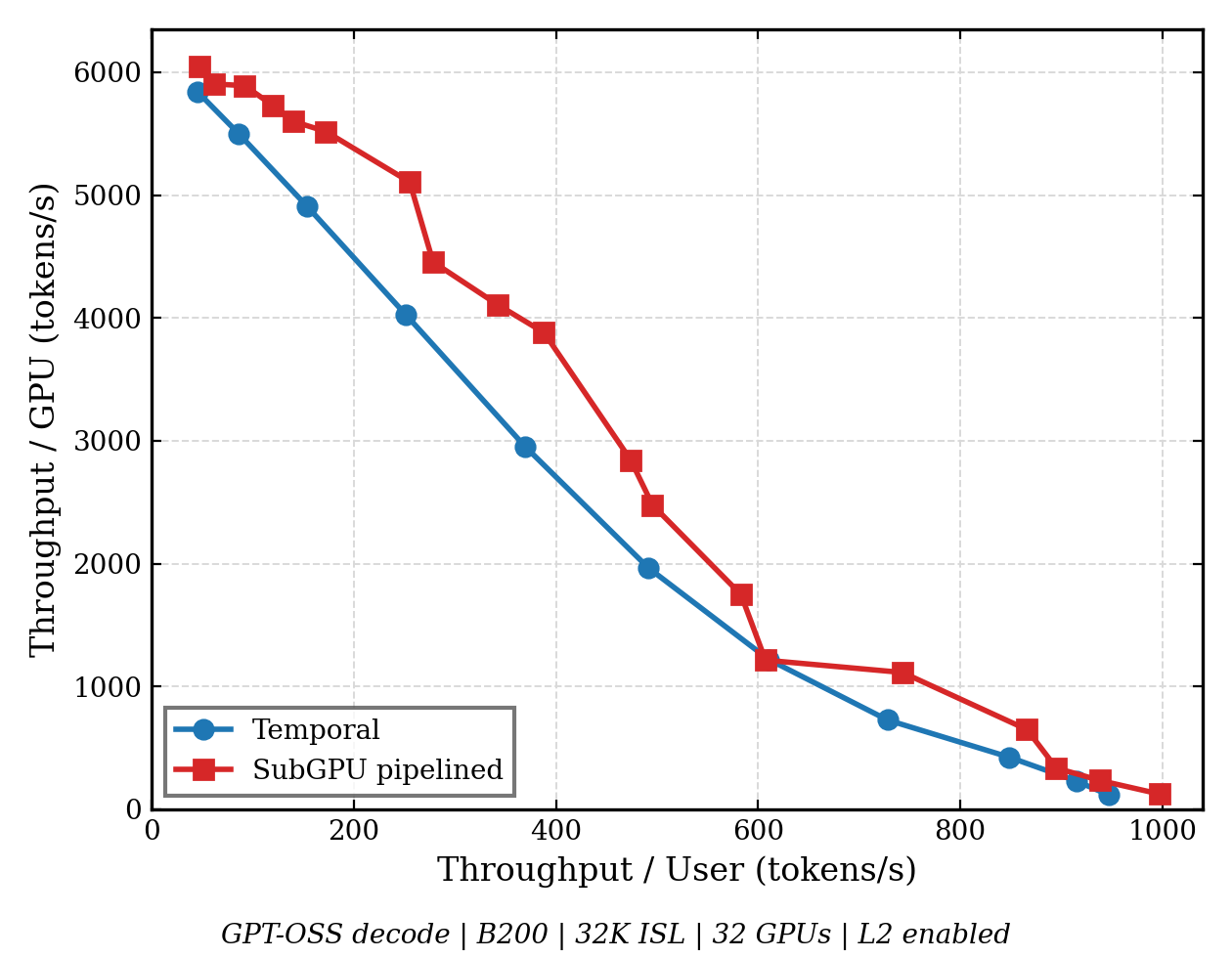}
    \caption{Modeled Pareto frontiers for GPT-OSS 120B decode on 32 B200 GPUs
    at input sequence length 32,768. The SubGPU curve pools the best retained
    configuration across stage counts.}
    \label{fig:pareto_base}
  \end{minipage}
  \hfill
  \begin{minipage}[t]{0.485\textwidth}
    \centering
    \includegraphics[width=\linewidth]{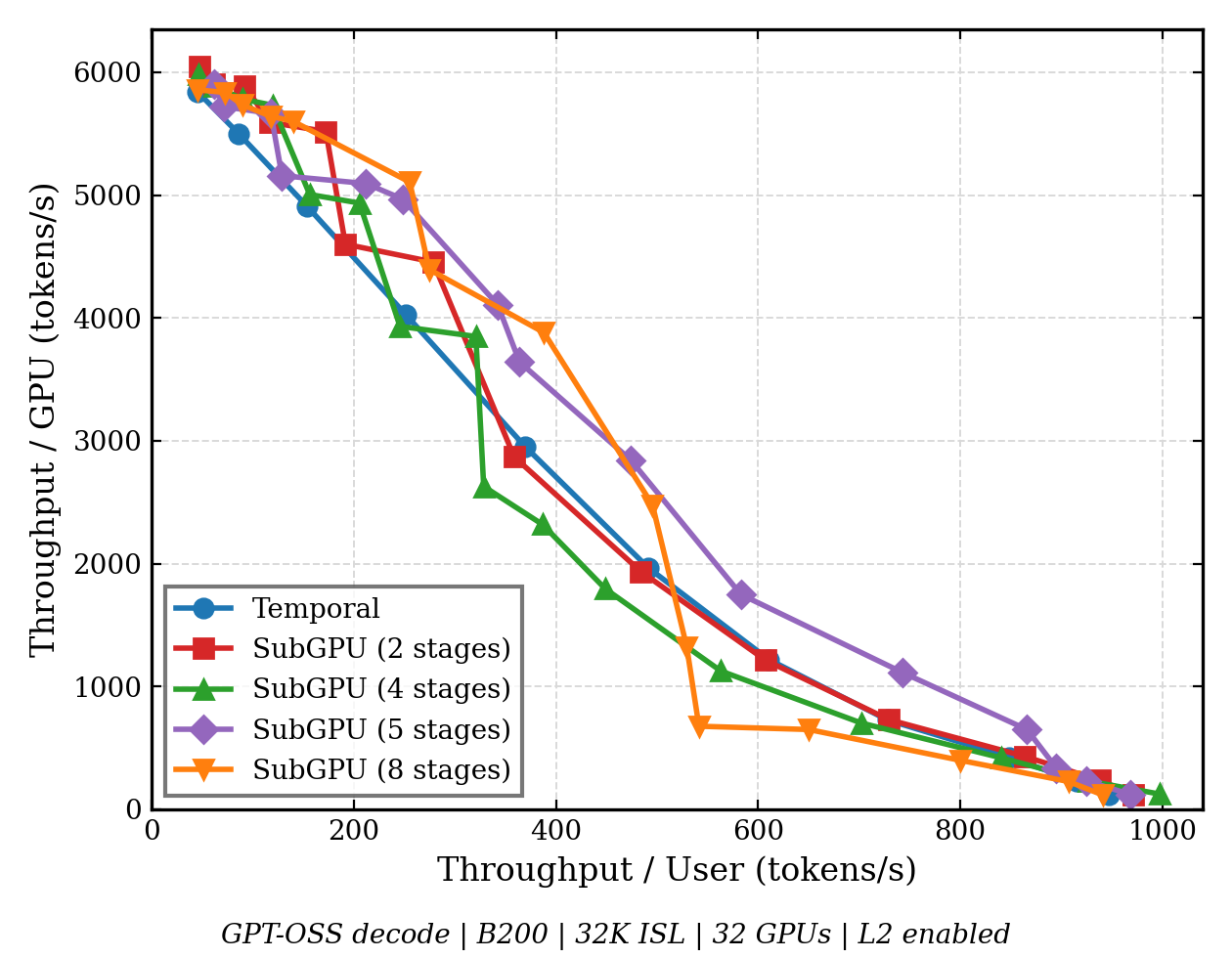}
    \caption{The same Pareto candidates separated by SubGPU stage count. Each
    stage is one physical SM partition; the temporal baseline uses the full SM
    pool.}
    \label{fig:pareto_pipeline}
  \end{minipage}
\end{figure*}

\section{Evaluation}
\label{sec:evaluation}

\subsection{Evaluation Methodology}
\label{sec:eval-method}

We evaluate SubGPU Pipelining in two complementary ways on the same
workload: GPT-OSS 120B~\cite{openai2025gptoss} decode at input sequence
length 32,768 and KV history 32,768, with 32-GPU B200 deployment settings: TP8/DP4 for attention and EP32 for MoE. Section~\ref{sec:simulation-results}
uses the analytical shared-resource simulator to search the full design space
and map the per-user/per-GPU throughput Pareto frontier;
Section~\ref{sec:hardware-results} implements one pipeline design on a B200
GPU with native TensorRT-LLM~\cite{nvidia2026tensorrtllm} kernels and measures its latency and resource
utilization.

In both evaluations, we compare:
\begin{itemize}[noitemsep,leftmargin=*]
  \item \textbf{Temporal}: conventional kernel-by-kernel execution, with each
  kernel eligible to use the full SM pool.
  \item \textbf{SubGPU}: the forward-pass graph is cut into contiguous
  segments, where each segment is bound to a stage with a static SM allocation, and
  microbatches are streamed through the stages.
\end{itemize}
The two evaluations serve different purposes. The simulation explores the
design space: each point on its frontier may use a different sharding, batch
size, microbatch size, graph segmentation, stage count, and SM allocation, so it
shows where SubGPU helps and which pipeline structures to build. The hardware
experiment runs the pipeline designs on real hardware at global batch 2,048
and reports the best-performing one in detail. It shows that the design can be realized
with existing CUDA mechanisms and kernels, and measures the latency and
utilization changes bring by SubGPU Pipelining.

% Previous version:
% Both evaluations use various pipeline designs that we developed for GPT-OSS's
% alternating attention pattern. GPT-OSS has
% repeating two-layer sliding/full-attention pattern rather than around one
% transformer layer in isolation. Sliding-window attention is substantially
% shorter than full attention, so placing both attention operators in one stage
% leaves that stage imbalanced across successive layers. Our designs therefore
% span the two-layer macro-period. We consider several designs, including
% additional stages spanning two layers and a stage dedicated to sliding-window
% attention. The simplest is the ``Macro'' two-stage design: one stage handles
% only full-attention FMHA, while the other handles all remaining operations,
% including sliding-window attention, across the two-layer macro-period.
Both evaluations use pipeline designs that we developed for GPT-OSS, whose
layers alternate between sliding-window and full attention. Sliding-window attention is substantially
shorter than full attention, so placing both attention operators in one stage
leaves that stage imbalanced across successive layers.
Therefore, we explore designs over a two-layer macro-period rather than a
single layer. 
We consider several such designs, including ones with additional
stages spanning two layers, and ones with a stage dedicated to sliding-window
attention. The simplest is the ``Macro'' two-stage design: one stage runs only
full-attention FMHA, and the other runs all remaining operators of the 
two-layer macro-period, including sliding-window attention. We evaluate these designs in both simulation and hardware.

% Previous evaluation scope and baselines subsection:
% \subsection{Evaluation Scope and Baselines}
%
% Previous evaluation framing:
% We use the analytical shared-resource simulator from \ref{sec:search} to estimate forward-pass makespan for temporal and SubGPU candidates, retaining the best found by the search. We stress that the curves are not meant to reflect or predict real model execution, rather to examine the benefits of spatial execution against our set of assumptions.

% The reported simulation sets kernel-launch overhead to zero for both execution models while retaining communication setup latency. This isolates the opportunity from spatial resource allocation and concurrent execution; the simulated improvement should not be attributed to launch-overhead elimination.

% We use the analytical shared-resource simulator from \ref{sec:search} to estimate forward-pass makespan for temporal and SubGPU candidates, retaining the best found by the search. The curves are not meant to predict measured execution, but rather to examine the benefits of spatial execution under our stated assumptions. We compare:
% \begin{itemize}[noitemsep,leftmargin=*]
%   \item \textbf{Temporal}: conventional kernel-by-kernel execution, with each kernel eligible to use the full SM pool.
%   \item \textbf{SubGPU}: contiguous graph cuts, boundary-reuse choice, stage count, static per-stage SM allocations, and microbatch size.
% \end{itemize}

\subsection{Analytical Simulation}
\label{sec:simulation-results}

\subsubsection{Simulation Setup}

For GPT-OSS 120B decode on 32 B200 GPUs, the simulator of
Section~\ref{sec:search} is used to estimate the makespan of one complete forward
pass for each temporal and SubGPU candidate, and we retain the best
configurations found by the search. The simulator unrolls all layers into the
forward-pass DAG, represents collectives as explicit tasks, and carries
shared-resource and L2-residency state across layer boundaries. We report the Pareto frontier of per-GPU throughput versus
per-user throughput, where per-user throughput serves as an inverse measure of
latency. The curves examine the benefit of spatial execution under our stated
modeling assumptions and are not meant to predict measured execution.

\subsubsection{Simulated Pareto Opportunity}

% Previous Figure 4:
% \begin{figure}[t]
%   \centering
%   \includegraphics[width=\linewidth]{figures/pareto_qwen3_r100_s131072_g64_1_baseline_pooled.pdf}
%   \caption{Modeled Pareto frontiers for temporal and SubGPU execution. Each point is the best retained configuration at that throughput tradeoff under the assumptions described in Sections~\ref{sec:search} and~\ref{sec:evaluation}.}
%   \label{fig:pareto_base_old}
% \end{figure}

% Figure~\ref{fig:pareto_base} compares the best retained frontiers. At matched per-GPU throughput in the range of approximately $1{,}000$--$3{,}000$ tokens/s/GPU, the modeled SubGPU configurations provide roughly $1.1$--$1.9\times$ higher per-user throughput than the temporal configurations. This region includes batches large enough to expose concurrent work while retaining heterogeneous compute, memory, and communication demands across the layer.

Figure~\ref{fig:pareto_base} compares the best retained frontiers. We compare
them in two ways, interpolating linearly between frontier points. At matched per-GPU throughput, SubGPU provides up to $2.2\times$ higher per-user
throughput than temporal execution. The gain is largest at the
throughput-oriented end of the frontier: it is $1.8$--$2.2\times$ above
5,000 tokens/s/GPU and $1.3$--$1.8\times$ between 3,000 and 5,000
tokens/s/GPU. At
matched per-user throughput, SubGPU provides $1.1$--$1.4\times$ higher per-GPU
throughput between 100 and 600 tokens/s/user and up to $1.9\times$ toward the
latency-oriented end.

The two frontiers approach parity at their extremes. At very small batches, there are too few independent microbatches to expose substantial overlap. At the opposite extreme, operators become more compute-bound and better utilize the temporal SM pool, reducing the capacity available for productive spatial co-scheduling. The modeled gains therefore arise from the joint configuration selected for the intermediate regime rather than from a universally faster segmentation or binding.

\begin{figure*}[t]
  \centering
  \includegraphics[width=\textwidth]{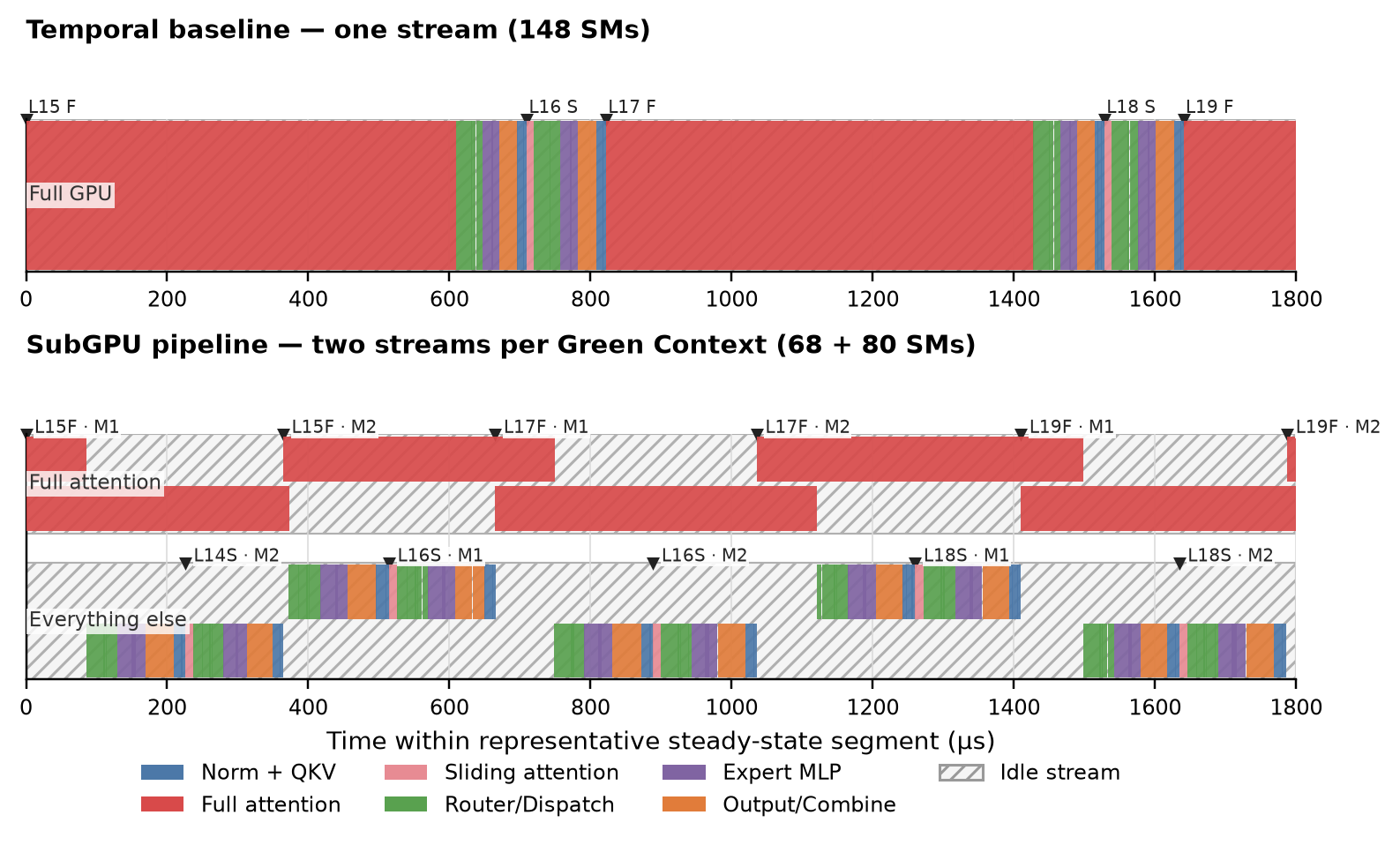}
  \caption{Kernel-level timeline for a representative 1.8-ms steady-state
  segment from the middle of the complete GPT-OSS replay at global batch 2048.
  The temporal baseline executes sliding-window attention, full attention, and
  the remaining operators on all 148 SMs. The SubGPU pipeline assigns
  full-attention FMHA to 68 SMs and everything else---including
  sliding-window attention and kernels from later layers---to 80 SMs. Each
  Green Context contains two CUDA stream rows. This is an instrumented representative
  replay; authoritative end-to-end timing comes from uninstrumented CUDA-event
  measurements.}
  \label{fig:trace_kernels}
\end{figure*}

\subsubsection{Effect of Stage Count}

Figure~\ref{fig:pareto_pipeline} groups the SubGPU candidates by stage
count. More
stages offer finer resource right-sizing and additional concurrency, but also
divide the SM pool more finely and require enough microbatch parallelism to
keep every stage productive. Since the first and last graph segments may
share a stage, a $D$-stage configuration contains either $D$ or $D+1$
contiguous segments. Stage count is therefore coupled to graph cuts, boundary
reuse, microbatch size, and per-stage allocation rather than being a fixed
architectural constant.

\subsection{Hardware Prototype on Blackwell}
\label{sec:hardware-results}

\subsubsection{Experimental Setup}

The silicon prototype of SubGPU Pipelining for GPT-OSS 120B is implemented with
TensorRT-LLM~\cite{nvidia2026tensorrtllm} on NVIDIA B200 GPU.
We used one B200 to emulate the per-rank workload of the 32-GPU deployment described in Section~\ref{sec:eval-method}. For this experiment, the global decode batch is chosen to be 2048 tokens. All computation kernels leverage the TensorRT-LLM kernels with version 1.3.0rc22.post1, while
communication are estimated using virtual kernels, performing the
corresponding attention all-reduce, MoE dispatch, and MoE combine endpoint work.
In our virtual communication model, each communication is floored at $3\,\mu\mathrm{s}$ latency
plus payload bytes divided by 900GB/s, representing the unidirectional B200 NVLink bandwidth.

\begin{figure*}[t]
  \centering
  \begin{subfigure}[t]{0.485\textwidth}
    \includegraphics[width=\linewidth]{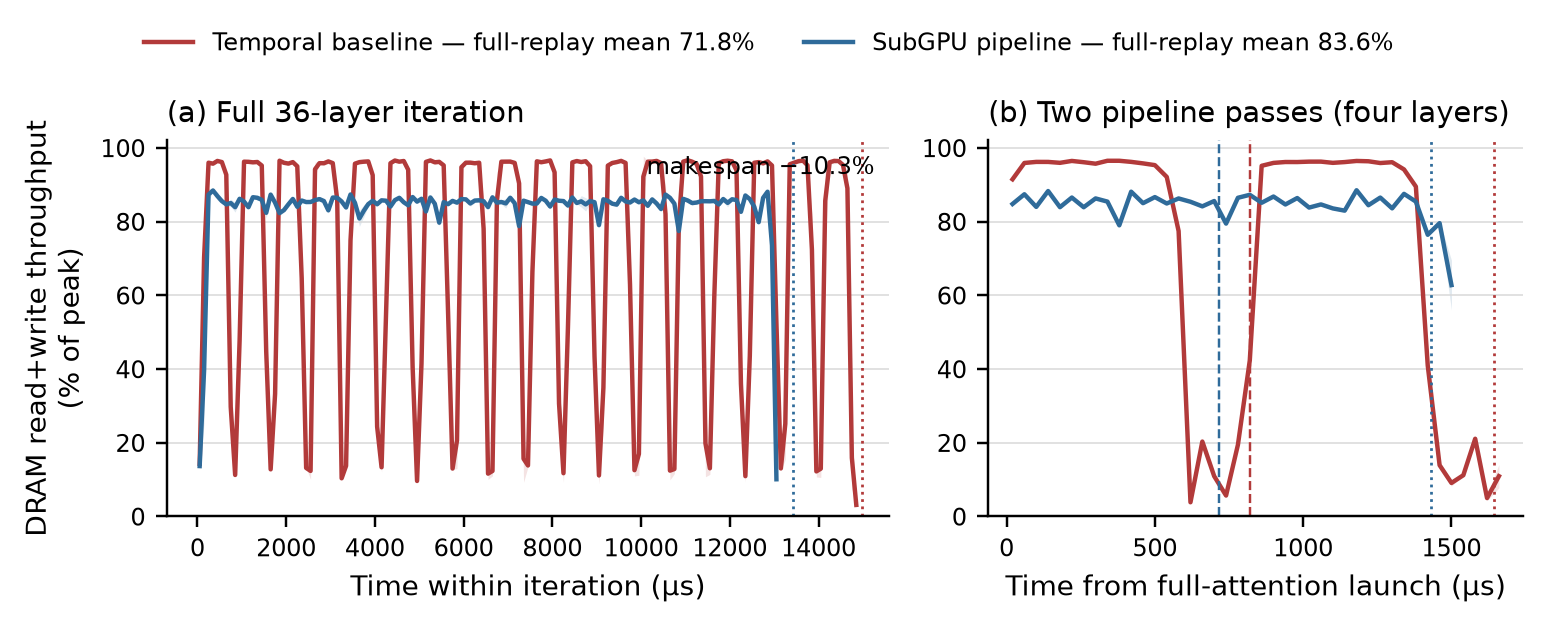}
    \caption{DRAM read-plus-write throughput. The SubGPU pipeline sustains
    83.6\% of peak on average, compared with 71.8\% for the temporal baseline.}
    \label{fig:trace_bw}
  \end{subfigure}
  \hfill
  \begin{subfigure}[t]{0.485\textwidth}
    \includegraphics[width=\linewidth]{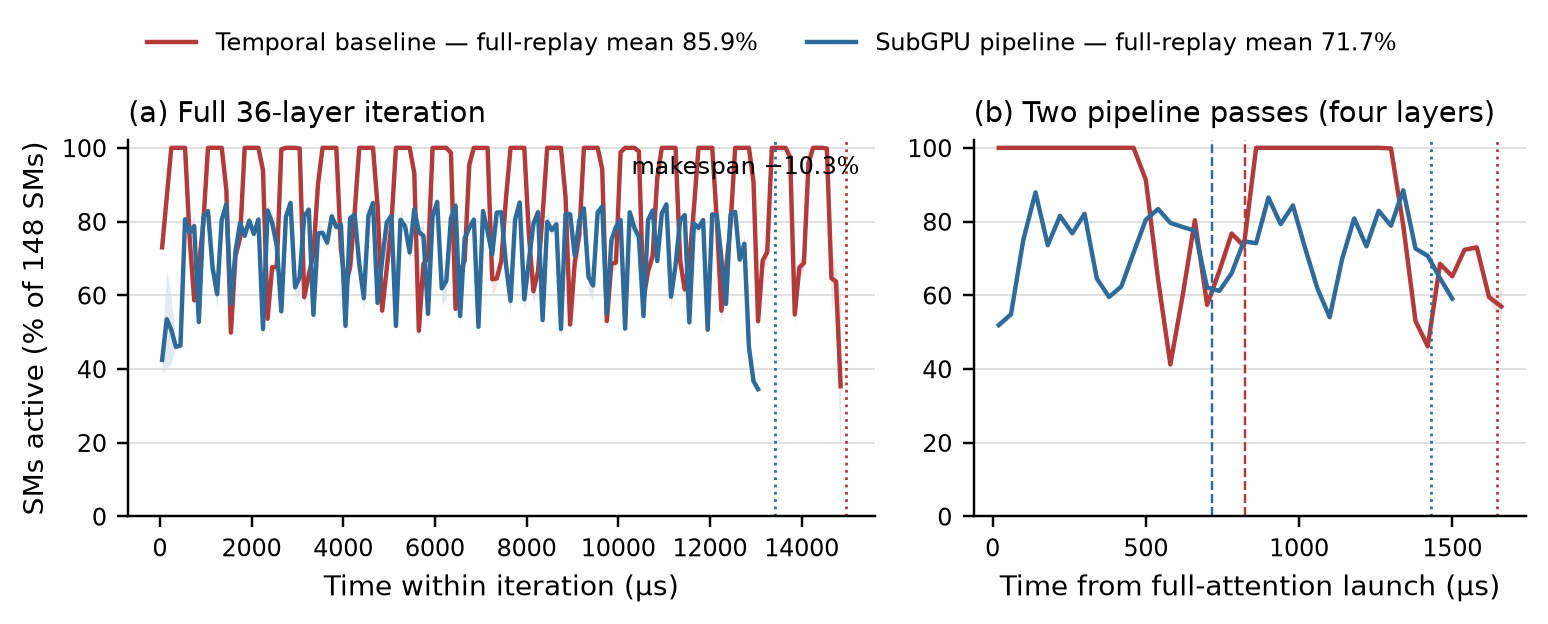}
    \caption{Fraction of the 148 SMs active. The SubGPU pipeline averages
    71.7\% and the temporal baseline averages 85.9\%}
    % TODO: ; the lower SubGPU value reflects partition-sized kernels rather than lower end-to-end throughput?
    \label{fig:trace_sm}
  \end{subfigure}
  \vspace{0.6em}

  {\small
  \begin{tabular}{@{}lrr@{}}
    \toprule
    Operator category & Temporal baseline (ms) & SubGPU pipeline (ms)\\
    \midrule
    Norm + QKV        & $0.556 \pm 0.001$ & $1.358 \pm 0.002$\\
    Full attention    & $10.957 \pm 0.008$ & $15.345 \pm 0.020$\\
    Sliding attention & $0.190 \pm 0.001$ & $0.426 \pm 0.002$\\
    Router/Dispatch   & $1.406 \pm 0.000$ & $3.473 \pm 0.010$\\
    Expert MLP        & $0.981 \pm 0.003$ & $3.059 \pm 0.012$\\
    Output/Combine    & $0.930 \pm 0.005$ & $2.789 \pm 0.017$\\
    \bottomrule
  \end{tabular}}
  \caption{Measured GPU utilization for one 36-layer GPT-OSS 120B decode
  iteration on an NVIDIA B200. Nsight Systems counters are sampled at 200\,kHz
  and aggregated into $100\,\mu\mathrm{s}$ bins for the full-iteration panels and
  $40\,\mu\mathrm{s}$ bins for the zoomed panels. Means are calculated from the unbinned 200-kHz samples. Left
  panels show the full iteration. Right panels zoom in on repeated sequences
  of two pipeline cycles, where each cycle spans one sliding-attention
  layer and one full-attention layer. 
  Dashed and dotted vertical lines indicate the mean durations of one and two
  pipeline cycles, respectively. The table gives the total kernel residence time for
  each category, reported as mean $\pm$ standard deviation across the four
  replays.}
  \label{fig:nsight_traces}
\end{figure*}

\subsubsection{Pipeline Design and Implementation}

We implement all purposed pipeline designs on
hardware, realizing every stage and SM partitions using CUDA Green Context. 
As actual kernel performance differs from the
SOL estimates used by the simulator, which shifts the relative durations of
the stages, for each of the pipeline designs, we tune the per-stage SM 
allocation to balances the execution of pipeline stages and minimize the measured
iteration latency, instead of directly reusing the SM allocations
from the simulator.

Across all designs and tuned allocations, the ``Macro'' two-stage pipeline design with
full-attention FMHA in a 68-SM Green Context and all other operators across two layers in an
80-SM Green Context achieves the lowest measured latency. We report this
configuration in the rest of this section. Each Green Context owns two
streams, and two equal microbatches flow through them. The temporal baseline
executes the same layer sequence as one batch on all 148 SMs. CUDA events and CUDA
Graphs enforce the producer/consumer and same-stream dependencies.
Figure~\ref{fig:trace_kernels} shows a representative steady-state segment
from the middle of the complete 36-layer replay.

\subsubsection{Latency and Resource Utilization Analysis}

After 20 warmups, three rounds of 100 uninstrumented CUDA Graph replays give a
mean 36-layer makespan of 14.970\,ms for the temporal baseline and 13.424\,ms
for the SubGPU pipeline, showing a 10.33\% latency reduction. 
In the corresponding 200-kHz Nsight captures, mean DRAM
read-plus-write throughput over four steady replays increases from 71.8\% to
83.6\% of peak. Mean active-SM throughput decreases from 85.9\% to 71.7\%:
partitioning uses fewer SMs for each individual kernel, but overlap raises
memory-system utilization and reduces end-to-end time.
Figure~\ref{fig:nsight_traces} shows these measured profiles.

\subsection{Comparing Simulation and Hardware Results}
\label{sec:sim-vs-hw}

The two evaluations agree in direction: at batch 2048, both find that SubGPU
Pipelining reduces decode latency relative to temporal execution, despite
differ in magnitude. At batch 2,048, the best simulated two-stage
configuration reduces makespan by 6.6\% (Section~\ref{sec:simulation-results}),
whereas the hardware measures a 10.33\% reduction.
We attribute part of this difference to optimistic per-kernel assumptions in the simulator. For
example, the model allows kernels to operate near 99\% of peak DRAM bandwidth,
whereas real execution incurs lower kernel efficiency and additional endpoint
overhead. These small costs accumulate along the serialized temporal baseline.
In the SubGPU pipeline, costs off the critical path can overlap with work in
the other context and be partially hidden. Therefore, the temporal baseline
can degrade more than the pipeline relative to their modeled performance,
producing a larger measured speedup. This tolerance of small per-kernel
inefficiencies is a second benefit of SubGPU Pipelining, in addition to
improving shared-resource utilization.

% Previous interpretation:
% The simulation and prototype answer different questions. The simulation
% estimates the opportunity available to the best configurations found by the
% staged search for Qwen3-235B-A22B-NVFP4 on modeled R100 deployments. The B200
% prototype demonstrates native-kernel, two-context spatial co-scheduling for
% alternating-attention GPT-OSS and reduces the measured iteration time by
% 10.33\%. We view this result as a proof of feasibility, with additional
% optimization still possible through different cuts, allocations, and kernels.
% The simulation results, by comparison, provide an aspirational performance
% envelope for other models and more aggressive optimization rather than an
% absolute prediction of the B200 experiment.

Taken together, the two results support each other. The simulation shows
where in the design space SubGPU Pipelining helps and which pipeline
structures are worth building. The hardware experiment shows that such a
structure can be realized with Green Contexts and unmodified kernels, and that
it delivers a measured 10.33\% latency reduction with higher DRAM utilization
on a real GPU. We view these results as a proof of feasibility, with additional
optimization still possible through different cuts, allocations, and kernels.

% ============================================================
\section{Realization, Extensions, and Composition}
\label{sec:realization}

\subsection{Implementation Considerations}

The logical model requires controlled partitioning of execution capacity, asynchronous progress of work on different stages, serialization of segments bound to the same stage, and dependency signaling. The timing model assumes that signaling and handoff overhead are negligible. Several CUDA mechanisms can provide parts of this functionality.

\paragraph{Green Contexts.}
CUDA Green Contexts~\cite{cuda124relnotes} bind streams to SM subsets and provide the resource isolation used by the current prototype.

\paragraph{CUDA events and graphs.}
The prototype uses events for dependency ordering and CUDA Graphs for repeatable low-overhead submission while retaining existing kernels.

\paragraph{Persistent kernels.}
A more specialized implementation could keep per-stage workers resident and feed them microbatches through work queues, potentially reducing launch overhead and enabling additional on-chip state reuse. This is an implementation option rather than a defining property of SubGPU and requires separate engineering and evaluation.

\subsection{Dynamic Reconfiguration}

Section~\ref{sec:model} notes that graph cuts, stage count, boundary reuse,
per-stage allocation, and microbatch size may vary between runtime regimes. A
runtime could select different configurations as KV-cache length, layer type,
routed expert load, or available batch parallelism changes. The appropriate
transition points and the cost of reconstructing or relaunching the underlying
execution resources are mechanism-specific. The simulator results and
prototype use static choices within each evaluated configuration.

\subsection{Weight Prefetch}
\label{sec:prefetch}

Prefetching the weights needed by segment $k$ for layer $L+1$ while other work from layer $L$ is active may reduce the first-use latency of the next layer. It is not free: the prefetch consumes HBM and L2 bandwidth, cache capacity, and possibly execution resources, and can therefore change the durations and overlap of every concurrently active segment. Temporal execution may also expose prefetch opportunities.  A thorough evaluation of prefetching is beyond the scope of this report, and is included in our on-going work on SubGPU Pipelining.

%\subsection{SubGPU Variants}
%
%A fully decomposed configuration places one operator in each graph segment. A $D$-stage configuration cuts the %graph into either $D$ segments or $D+1$ segments with boundary reuse, trading off resource specialization, %concurrency, and pipeline overhead. The prototype is a two-stage, three-segment configuration: its first and last %segments share the rest-of-layer stage, and its middle segment uses the FMHA stage. These are points in the same %design space rather than separate execution models. We can also extend the design space by considering graph %segments mapped to multiple static partitions as opposed to the single static partition we discuss here.

\subsection{Composition with Outer Parallelism}

SubGPU operates within the per-GPU work allocation produced by outer tensor, data, expert, context, or pipeline parallelism, although the ideal parallelism strategy may change when SubGPU is used. The outer strategy determines each GPU's operator graph, tensor shapes, communication, and batch; the SubGPU optimizer then cuts that graph and binds its segments to local stages. When pipeline parallelism is present, SubGPU forms an inner, intra-GPU schedule within each pipeline rank. Outer communication must appear as tasks or data edges in the augmented task DAG, and its shared-resource effects must be reflected in the Section~\ref{sec:search} duration model.

% ============================================================
\section{Discussion and Limitations}
\label{sec:discussion}

\paragraph{Applicability conditions.}
SubGPU relies on enough independent work to overlap work across stages. It is most promising when the batch exposes multiple microbatches, operators have complementary resource profiles, and some temporal operators reach their bandwidth or parallelism limit before using the full SM pool. The opportunity diminishes when the batch is too small to populate the pipeline or when large, compute-bound kernels already use the GPU efficiently.

Microbatching also changes weight-reuse behavior. If concurrently active
segment working sets do not remain in L2, repeated HBM traffic can offset the
benefit of spatial overlap. This limitation becomes more important
for larger models with larger weights and activations. Techniques such as
prefetching data into L2 may mitigate this overhead, but require further
evaluation.

\paragraph{Modeling limitations.}

The simulation results are modeled opportunities rather than formal upper
bounds. They use static within-configuration allocations and configured
per-kernel base overheads rather than modeling launch and synchronization
dynamics in detail. L2 residency is enabled for modeled weights and
activations, while KV-cache accesses remain streaming traffic that is not
admitted to the resident-state table. Their credibility depends on the
resource-sharing, cache-initialization, admission, eviction, communication, and
operator-scaling policies described in Section~\ref{sec:search}. Sensitivity to
these assumptions and calibration against measured kernels are necessary
before extrapolating the Pareto gains to an end-to-end deployment.

\paragraph{Implementation complexity.}
Any implementation must manage graph-cut and SM-partition granularity, dependency signaling, buffering, failures, and transitions between configurations. Persistent-kernel implementations add work-queue, SMEM-lifetime, and recovery concerns, while framework-level implementations must control launch and synchronization overhead. These costs are mechanism-specific and are not part of the logical definition.

\paragraph{Future work.}

The next steps are to validate the simulator and optimizer more extensively,
compare selected configurations with exhaustive search on small instances, and evaluate
end-to-end dense and MoE deployments across GPU generations. Dynamic
allocation, persistent-kernel realizations, weight prefetch, and longer-lived
on-chip weight persistency remain extensions to be evaluated with their
contention and transition costs included.

% ============================================================
\section{Conclusion}
\label{sec:conclusion}

% Previous conclusion:
% We presented SubGPU Pipelining for fully utilizing GPU compute and memory throughput. Its defining property is not a particular CUDA primitive, but a schedule that binds ordered contiguous graph segments to controlled SM-backed stages and overlaps work on different stages while preserving graph dependencies and stage exclusivity. Under the assumptions of the analytical model, the best retained SubGPU configurations provide roughly $1.1$--$1.9\times$ higher per-user throughput than temporal execution at matched per-GPU throughput in the latency-sensitive regime. A separate two-stage Blackwell prototype maps three graph segments to two Green-Context-backed SM pools, with its first and last segments sharing one stage; it raises mean DRAM read-plus-write throughput from 69\% to 78\% of peak and reduces the mean layer period by 8.8\% in its evaluated configuration. Together, these results identify a modeled opportunity and demonstrate that its core spatial co-scheduling mechanism is realizable, while leaving broader model validation, end-to-end evaluation, and more specialized implementations to future work.

We presented SubGPU Pipelining, a technique for improving GPU compute and
memory utilization. SubGPU Pipelining partitions the
forward-pass graph into contiguous segments, binds each segment to a dedicated
pool of SMs, and streams microbatches through these pools so that operators
with different resource demands run concurrently. The approach is defined by
this schedule rather than by any particular CUDA primitive.
We evaluated SubGPU Pipelining on GPT-OSS 120B in a 32-GPU B200 deployment setting.
In simulation, SubGPU configurations outperform temporal execution across most
of the per-user/per-GPU throughput Pareto frontier, providing up to
$2.2\times$ higher per-user throughput at matched per-GPU throughput and up to
$1.9\times$ higher per-GPU throughput at matched per-user throughput. On real
hardware, a
two-stage implementation built from CUDA Green Contexts and existing
TensorRT-LLM kernels raises mean DRAM throughput from 71.8\% to 83.6\% of peak
and reduces the iteration time by 10.33\%. This result shows that SubGPU
Pipelining is physically feasible on current GPUs and delivers a measurable
benefit without custom kernels. Validating additional configurations and exploring more
specialized implementations remain future work.

% ============================================================
\section*{Acknowledgments}

We thank various teams within NVIDIA for their support. This includes Jeremy Wilke, Wonchan Lee, Anurag Gupta, Haoming Lu, Yuanqing Zhao, Jatin Mitra, Ankit More, Brian Pharris, Angshuman Parashar, Vikram Sharma Mailthody, Neal Crago, Guilleme Sole, Andrew Foote.

\subsubsection*{Declaration of Generative AI and AI-assisted technologies in the writing process}

During the preparation of this work, the authors used generative AI to improve the clarity and fluency of the English text as well as some scripts. After using these tools, the authors reviewed and edited the content as needed and take full responsibility for the publication's content.
% ============================================================
\begingroup
\small
\bibliographystyle{plain}
\bibliography{subwave_pipelining}
\endgroup

\end{document}